\documentclass[epj]{svjour}
\usepackage{graphics}
\usepackage{graphicx}
\usepackage{physics}
\usepackage{hyperref}
\usepackage{url}
\begin{document}
\title{Constructing Lifshitz spaces using the Ricci flow}
\author{R. Cartas-Fuentevilla\inst{1},  A. Herrera-Aguilar\inst{1} \and J. A. 
Herrera-Mendoza \inst{1}% etc
}                     % Do not remove
\institute{Instituto de F\'{i}sica-LRT, Benem\'{e}rita Universidad Aut\'{o}noma 
de Puebla, 72550, Puebla, Mexico}
\date{Received: date / Revised version: date}
\abstract{
In this work we make use of the Ricci flow equations to show that, by starting 
from a general ansatz for the metric, we can construct two kinds of Lifshitz 
spaces in which: (a) the critical exponent coincides with the spatial dimension of 
the spacetime and therefore adopts discrete values, and (b) the critical exponent is 
continuous and arbitrary. These results show that Lifshitz spaces are exact solutions 
to the Ricci flow equations. Moreover, we found that the Ricci flow evolves towards a 
single fixed point for both cases which coincides with the flat spacetime.
%
%\PACS{
%   {PACS-key}{}   \and
    % {PACS-key}{discribing text of that key}
		%} % end of PACS codes
} %end of abstract
\maketitle
\section{Introduction}
The concept of Ricci flow was originally introduced by Richard Hamilton in 1982 
\cite{Hamilton} as a part of a developed program to resolve the well-known 
Poincar\'{e} conjecture or -more generally- the Thurston geometrization 
conjecture for closed 3-manifolds. The Ricci flow describes a mathematical 
treatment that allows us to continuously deform a Riemannian manifold, such 
deformation is realized by means of a partial differential equation that behaves 
like a kind of diffusion heat equation for the metric. An equivalent formulation 
was introduced by Dennis DeTurck \cite{Turck} through a family of diffeomorphisms 
along the flow, giving rise to the Hamilton-DeTurck Ricci flow.
		
In physics, this tool has gained some attention, in part thanks to Perelman's 
wonderful work \cite{Perelman1,Perelman2} proving the geometrization conjecture  
almost a century after the Poincar\'{e} conjecture was proposed. The most 
obvious applications of Ricci's flow in physics can be realized within the 
framework of general relativity, since it is by its nature a theory about the 
geometry of spacetime. In this context, the Ricci flow has been used to model 
the numerical evolution of wormholes and black holes 
\cite{Husain}-\cite{Wiseman}. The evolution of Morris-Thorne wormhole geometries 
and bubble geometries using numerical methods was studied in \cite{Husain}. The 
question on how the Arnowitt-Deser-Misner (ADM) mass change under the Ricci flow 
was analyzed in \cite{Dai}. Particularly, the study of the ADM mass of an 
asymptotic locally euclidean (ALE) space along the Ricci flow was approached. 
Moreover, the authors showed that the ALE property  is preserved along the Ricci 
flow and, for a three-dimensional space, the mass is an invariant under the 
Ricci flow. In contrast with these results, the decay of the ADM mass of an 
asymptotically hyperbolic manifold of dimension $d\ge 3$ has been shown  in 
\cite{Woolgar2012}. Additionally, starting from the decay formula for the 
asymptotically hyperbolic mass under the curvature-normalized flow, an heuristic 
derivation on the invariance of the ADM mass for an asymptotically flat Ricci 
flow was given.  
		
From Perelman's gradient formulation for the Ricci flow, a possible connection 
between Perelman entropy and Bekenstein-Hawking entropy (geometric entropy) that 
arises from black hole thermodynamics was investigated in \cite{Samuel2007}. 
However, by studying the corresponding fixed points for the flow, no connection 
was found between these two entropies. Nevertheless, the authors proposed a new 
flow that apparently admits such connection and may have applications in black 
holes physics by suggesting new approaches to Penrose inequality. In 
\cite{Samuel2008} the Ricci flow  was implemented to describe the evolution of 
the area and Hawking mass of a 2-dimensional closed manifold. To get a physical 
significance of the results there are two facts from general relativity that 
were considered; the area of event horizons is related with the black hole 
entropy and,  additionally, the fact that the Hawking mass of an asymptotic 
round 2-sphere is just the ADM mass. As a main result, an inequality which 
relates the evolution of the area of a closed surface and the Hawking mass was 
derived.
		
In \cite{Wiseman} the Ricci flow was applied to the study of 4-dimensional 
Euclidean gravity with boundary $S^{1}\times S^{2}$ which represents the 
canonical ensemble for gravity in a spherical box. It was found that at high 
temperatures the action has three saddle points: hot flat space, and large and 
small black holes. The small black hole is unstable under the Ricci flow since 
it has a Perry-Yaffe-type negative mode, while the other two are stable under 
the Ricci flow.

A review on how the Ricci flow arises in the renormalization group (RG) of 
non-linear sigma models was presented in \cite{Woolgar2008}. Moreover, 
in this work a 
discussion of the behaviour of the mass under the Ricci flow by using a 
two-dimensional, rotationally symmetric, Ricci soliton was given. Additionally, 
the connection between the holographic renormalization group flow and the Ricci flow 
has been revealed in \cite{jackson,kiritsis}; particularly, in \cite{jackson} 
the authors introduced the Hamilton-Jacobi formalism to derive the RG flow 
equations and to show that, for the case of dual AdS/QFT theories, they are described 
by the Ricci flow. Another physical theory in which the 
Ricci flow naturally emerges is the well-known Ho\v{r}ava-Lifshitz gravity 
\cite{horava,nishioka};  the original Ricci flow arises as a limiting 
case of a more general RG flow obtained by applying the Hamilton-Jacobi 
equation to the Ho\v{r}ava-Lifshitz gravity theory and setting some parameters 
\cite{nishioka}. The maximum principle was 
used in \cite{Figueras} to show that  a Ricci soliton does not exist in 
Riemannian manifolds with boundaries such as those static Lorentzian spacetimes 
which are asymptotically flat, Kaluza Klein, locally AdS or have extremal 
horizons.

Recently, the study of Ricci flows for maximally symmetric manifolds with 
constant curvature was performed in \cite{Arturo}. In this work it was found 
that spacetimes with positive constant curvature (de Sitter spaces) evolve into 
spacetimes with negative constant curvature (Anti-de Sitter spaces) under the 
Ricci flow through a singularity in the curvature, evolving further into 
Minkowski spacetime. There is another family of spacetimes with negative 
constant curvature that has a group of symmetries distinct from that of Anti-de 
Sitter space (which is a maximally symmetric space). The metric of these 
manifolds is invariant under anisotropic scalings of time and space; these 
spaces are called Lifshitz spacetimes and are given by the metric:
\begin{equation} \label{metricL}
ds^{2} = \ell^{2} \left( -r^{\pm 2z} dt^{2} + r^{\pm 2} dx^{2}_{i} + 
\frac{dr^{2}}{r^{2}}\right), \qquad i =1,2,\ldots,D,
\end{equation}
where $z$ is a continuous parameter known as the critical exponent. This 
metric has a symmetry group denoted by ${\bf{Lif}}_{D}(\bf{z})$ 
\cite{Taylor,Viri}, and is invariant under translations and spatial rotations 
($H$, $P^i$, $L^{ij}$) 
\begin{equation}\label{gs1}
\begin{split}
&H:  t \longrightarrow t'=t+a;	\\
&P^{i}: x^{i} \longrightarrow x^{'i}=x^{i}+a^{i}; \\   
&L^{ij}: x^{i} \longrightarrow x^{'i}=L^{i}_{j}x^{j},
\end{split}
\end{equation} 
and the non relativistic scaling symmetry $D_{z}$
\begin{equation}\label{gs2}
\begin{split}
D_{z}:\ \ 
 & r \longrightarrow r' =\lambda^{\pm 1}r, \\
  & t \longrightarrow t' =\lambda^{\mp z}t, \\ 
& x^{i} \longrightarrow x^{'i} =\lambda^{\mp 1} x^{i}. 
\end{split}
\end{equation}
The main purpose of this work is to introduce the Hamilton-DeTurck Ricci 
flow to study the construction of spacetime geometries with particular physical 
interest focused on those that admit Lifshitz symmetries.
We already know that Lifshitz spaces arise as solutions to the Einstein 
field equations with several non-trivial stress energy tensors that depend 
on a given theory (see, for instance, \cite{Taylor,Kachru,HLS}). 
On the other hand, the Ricci flow is a purely geometric equation that involves an external 
parameter that can be linked to a physical quantity; this is a very interesting fact since our results imply 
that one can construct Lifshitz spacetimes in a purely geometrical way, dispensing with the physical fields, 
for a wide family of field theories that support such metrics. Another important aspect of the study of 
Lifshitz spaces resides on the 
fact that they play a very important role in the description of the holographic 
quantum systems within the so-called  Gravity (Lifshitz)/Condensed Matter 
Theory correspondence \cite{Taylor}. A better study of the intrinsic 
properties of the Lifshitz spaces will lead us to a deeper understanding of the 
holographic correspondence as well as to interesting applications to quantum 
physical systems through Lifshitz holography.

After getting exact solutions to the Ricci flow equations, we show that by 
means of an appropriate transformation of coordinates on the obtained metric, 
we get a family of Lifshitz spaces with discrete and continuous critical 
exponent. These metrics tends to a fixed point that corresponds to  flat 
Minkowski spacetime as the flow evolves. 
		
		The structure of this work is as follows, in section \ref{s2} we introduced 
the Hamilton-DeTurck formulation of the Ricci flow. Then, we start with a quite 
general spacetime geometry that depends on the energy coordinate $r$ and the 
Ricci parameter $\lambda$, and evolves under the Hamilton-DeTurck Ricci flow. 
This evolution will render a set of nonlinear coupled partial differential 
equations. To solve this system, we assume an ansatz on the DeTurck vector field 
in such way that the system becomes consistent under the Hamilton-DeTurck Ricci 
flow. We further find exact solutions to the flow equations that can be recast 
as a Lifshitz metrics; here we also reveal that there exist one fixed point 
along  the flow that reproduces the flat spacetime. In section \ref{s3} we present some 
concluding remarks and set a discussion about the fixed point of the geometry 
found for the Lifshitz metric constructed in this work.
		
\section{Lifshitz space construction}\label{s2} 
Hamilton's original formulation for the Ricci flow is defined by the equation \cite{Topping}: 
\begin{equation}\label{eq1} \partial_{\lambda} g_{\mu \nu } = - 2 R_{\mu \nu},  \ \ \ \ \  g(0) = g_{0}, 
\end{equation} 
where $R_{\mu \nu}$ is the Ricci tensor constructed from the metric $g_{\mu \nu}$, which additionally depends on the flow parameter $\lambda$ and $g_0$ is the initial condition on the metric when $\lambda=0$. This equation describes the evolution of the geometry under the flow parameter that is external to the spacetime in question. Since (\ref{eq1}) is diffeomorphism invariant, it is a degenerate nonlinear PDE and therefore we say that it is weakly parabolic \cite{benneth}. Nevertheless, DeTurck showed that it is possible to modify the original formulation of the Ricci flow and obtain a strongly coupled parabolic nonlinear PDE  \cite{Turck,Topping}. This new formulation is named the Hamilton-DeTurck Ricci flow and it is defined by: 
\begin{equation}
\label{eq2} 
\partial_{\lambda} g_{\mu \nu } = - 2 R_{\mu \nu} + \nabla_{\mu} V_{\nu}+ \nabla_{\nu} V_{\mu}, \end{equation} 
here $V_{\mu}$ is the DeTurck vector field that generates diffeomorphisms along the flow.
		
We shall begin by proposing a general geometry given by
 \begin{equation}\label{eq3} d{s^2} = l^2 \qty[- f_{1}(\lambda,r) d{t^2}+\frac{1}{r^2} d{r^2}+ f_{3}(\lambda,r) d{x_{i}}d{x^{i}}], \ \ \ \ \   i = 1,2,\ldots D, \end{equation} 
where $(t,x_{i})$ stand for time and spatial coordinates, respectively, while $r$ denotes an extra coordinate that has the physical meaning of energy within the holographic correspondence framework, and $f_{1}(\lambda,r)$ and $f_{3}(\lambda,r)$ are arbitrary functions of $r$ and the Ricci flow parameter $\lambda$\footnote{We could in principle, add one more arbitrary function $f_{2} (\lambda,r)$ in the $g_{rr}$ component of this metric. Here we set this function to unity for the sake of simplicity.}. We wish to construct solutions to the Hamilton-DeTurck Ricci flow (\ref{eq2}). First of all, we need to compute the non-zero Christoffel symbols associated with the metric (\ref{eq3})
 \begin{equation}\label{eq4} 
 \begin{split} \varGamma^{t}_{\ t r} &= \frac{1}{2} \frac{f_{1}'(\lambda,r)}{f_{1}(\lambda,r)},\\ \varGamma^{r}_{\ t t} &= \frac{1}{2} r^{2} f_{1}'(\lambda,r),\\ \varGamma^{r}_{\ i j} &= -\frac{1}{2} r^{2} f_{3}'(\lambda,r)\delta_{i j}, \end{split} \hspace{4cm} \begin{split} \varGamma^{r}_{\ r r} &= -\frac{1}{r},\\ \varGamma^{i}_{\ j r} &= \frac{1}{2} \frac{f_{3}'(\lambda,r)}{f_{3}(\lambda,r)}\delta^{i}_{j}.\\ 
 \end{split} 
 \end{equation}
With the aid of these symbols, we now compute the non-zero components of the Ricci tensor: \begin{equation}\label{eq5} \begin{split} R_{tt} &= \frac{r^2}{2} f_{1}'' -\frac{r^2}{4} \frac{{f_{1}'}^2}{f_{1}}+\frac{r}{2} f_{1}'+\frac{D r^2}{4} \frac{f_{1}'f_{3}'}{ f_{3}},\\ R_{rr} &= -\frac{1}{2} \frac{f_{1}''}{f_{1}} -\frac{D}{2} \frac{f_{3}''}{f_{3}} +\frac{1}{4} \frac{{f_{1}'}^2}{f_{1}^2} + \frac{D}{4} \frac{{f_{3}'}^2}{f_{3}^2}-\frac{1}{2 r} \frac{f_{1}'}{f_{1}}-\frac{D}{2 r}\frac{f_{3}'}{f_{3}},\\ R_{i j} &= \qty[-\frac{r^2 f_{3}''}{2}+\qty(2-D)r^2 \frac{ {f_{3}'}^2}{4 f_{3}}-\frac{r}{2} f_{3}'-r^2 \frac{f_{1}' f_{3}'}{4 f_{1}}] \delta_{ij}. \end{split} \end{equation}
		
We are now in position to consider the Hamilton-DeTurck Ricci flow (\ref{eq2}) and find a system of $D+2$ nonlinear coupled partial differential equations (only three of them are independent) that describes the flow. Here we shall invoke our first ansatz by assuming that the DeTurck vector field has only  $r$ component, that is $V_{\mu } =\qty(0, V_{r} (\lambda,r), 0, \ldots,0)$. With this fact in mind, the independent flow equations read 
\begin{align} 
l^2 \dot{f}_{1} &= r^2 f_{1}''-\frac{r^2}{2} \frac{{f_{1}'}^2}{f_{1}}+r f_{1}'+\frac{D r^2}{2} \frac{f_{1}' f_{3}'}{f_{3}}+r^2 f_{1}' V_{r} \label{eq6},\\ 
l^2 \dot{f}_{3} &= r^2 f_{3}''- \frac{(2-D)}{2}r^2 \frac{{f_{3}'}^2}{f_{3}} + r f_{3}' + \frac{r^2}{2} \frac{f_{1}'f_{3}'}{f_{1}} +r^2 f_{3}' V_{r} \label{eq7}, \\ 
0 &= \frac{f_{1}''}{f_{1}}-\frac{1}{2}\frac{{f_{1}'}^2}{{f_{1}}^2} +\frac{1}{r} \frac{f_{1}'}{f_{1}}+D\qty[\frac{f_{3}''}{f_{3}}-\frac{1}{2} \frac{{f_{3}'}^2}{{f_{3}}^2}+\frac{1}{r} \frac{f_{3}'}{f_{3}}]+2 V_{r}'+\frac{2}{r} V_{r} \label{eq8}, 
\end{align} 
where the dots represent derivatives with respect to the flow parameter $\lambda$ and the primes are derivatives with respect to the coordinate $r$. For the sake of simplicity we shall perform the following transformation $f_{j}(\lambda,r) = \text{e}^{u_{j}(\lambda,r)}$, where $j=1, 3$, rendering the following system
\begin{align} l^2 \dot{u}_{1} &= r^2 u_{1}''+\frac{r^2}{2}{u_{1}'}^2+r u_{1}'+\frac{D}{2}r^2 u_{1}' u_{3}'+r^2 u_{1}' V_{r} \label{eq9},\\ 
l^2 \dot{u}_{3} &= r^2 u_{3}''+ \frac{D}{2} r^2 {u_{3}'}^2 + r u_{3}' + \frac{r^2}{2} u_{1}'u_{3}' +r^2 u_{3}' V_{r} \label{eq10}, \\ 
0 &= u_{1}''+\frac{1}{2}{u_{1}'}^2 +\frac{1}{r} u_{1}'+D\qty[u_{3}''+\frac{1}{2}{u_{3}'}^2+\frac{1}{r} u_{3}']+2 V_{r}'+\frac{2}{r} V_{r} \label{eq11}. \end{align} 

By adding the equations (\ref{eq9}) and (\ref{eq10}) we can define a new function 
\begin{equation}\label{eqw} w(\lambda,r) = u_{1}(\lambda, r)+ D u_{3}(\lambda,r) \end{equation} 
such that the sum of these equations gives 
\begin{equation}\label{eq12} 
l^2 \dot{w} = r^2 w''+\frac{r^2}{2} {w'}^2+r w'+r^2 w' V_{r}, 
\end{equation} 
%% 
%%%%%%%% SUBSECTION %%%%%%%%%
\subsection{Solutions to the flow with discrete critical exponent}  %%%%%%%%%%%
%%%%%%%%
By taking the definition of $w(\lambda,r)$ we can express (\ref{eq11}) as follows 
\begin{equation}\label{eq13} \frac{D (D+1)}{2} {u_{3}'}^2 - D w' u_{3}'+ w''+\frac{1}{2} {w'}^2 +\frac{1}{r} w'+ 2V_{r}'+ \frac{2}{r} V_{r}=0. \end{equation} 
Since $V_{r}$ is an arbitrary function of $r$ and $\lambda$, we choose our second ansatz such that \begin{equation}\label{anz2} w''+\frac{1}{2} {w'}^2 +\frac{1}{r} w'+ 2V_{r}'+ \frac{2}{r} V_{r}= -\frac{{q(\lambda)}^2}{r^2}, \end{equation} where $q(\lambda)$ is an arbitrary function of $\lambda$. This gives us a system of PDE's that couples $w(\lambda,r)$ with $V_{r}(\lambda,r)$: 
\begin{align} l^2 \dot{w} &= r^2 w''+\frac{r^2}{2} {w'}^2+r w'+r^2 w' V_{r}, \label{eq14}\\ 0 &= r^2 w''+\frac{r^2}{2} {w'}^2 +r w'+ 2 r^2 V_{r}'+ 2 r V_{r}+{q(\lambda)}^2. \label{eq15} \end{align} 
 As a simple solution for this system we choose 
 \begin{equation}\label{eq17} 
 \begin{split} 
 V_{r}(\lambda,r) &= -\frac{1}{2} q(\lambda)^2 \frac{\ln(r)}{r} + \frac{h (\lambda)}{r},\\ w(\lambda,r) &=c= \text{const}. 
 \end{split} 
 \end{equation} 
where $h(\lambda)$ is one more arbitrary function of $\lambda$. Then, by using (\ref{eq13}) we can find solutions for $u_{3}$ and then for $u_{1}$, obtaining: 
\begin{align} 
u_{1} (\lambda,r) &= \pm D\sqrt{\frac{2}{D(D+1)}} q(\lambda) \ln r - D m(\lambda)+ \text{const.} \label{eq18},\\
u_{3} (\lambda,r) &= \pm \sqrt{\frac{2}{D(D+1)}} q(\lambda) \ln r+ m(\lambda) \label{eq19}, 
\end{align} 
where $m(\lambda)$ is an arbitrary function of $\lambda$. By substituting (\ref{eq18}) and (\ref{eq19}) into (\ref{eq9}) and (\ref{eq10}) we find some restrictions upon $g(\lambda)$, $h(\lambda)$ and $m(\lambda)$. By considering the solutions with either plus or minus sign before the square root of both (\ref{eq18}) and (\ref{eq19}) we obtain 
\begin{equation}\label{eq20} 
\begin{split} 
q(\lambda) &= \sqrt{\frac{l^2}{\lambda-\lambda_{0}}}, \\ h(\lambda) &= \mp\sqrt{\frac{2 D l^2}{(D+1)(\lambda-\lambda_{0})}},\\ m(\lambda) &= 0. 
\end{split} 
\end{equation}
These restrictions ensure that the system is solved in a consistent way. Therefore, the solutions for $u_{1}$ and $u_{3}$ are given by: 
\begin{equation}\label{eq21}
 u_{1}(\lambda,r) = \pm D\sqrt{\frac{2 l^2}{D(D+1)(\lambda-\lambda_0)}} \ln r+c, \qquad u_{3}(\lambda,r) = \pm \sqrt{\frac{2 l^2}{D(D+1)(\lambda-\lambda_0)}} \ln r, 
 \end{equation} 
since $c$ is a constant, it can be absorbed into the coordinates and it does not matter if we choose it as zero. Therefore 
\begin{equation}\label{eq22} f_{1}(\lambda,r) = r^{\pm D\alpha(\lambda,D)}, \qquad  f_{3}(\lambda,r) = r^{\pm \alpha(\lambda,D)}, 
\end{equation} 
where the involved parameter $\alpha(\lambda,D)$ reads 
\begin{equation}\label{eq23} \alpha(\lambda,D) = \sqrt{\frac{2 l^2}{D(D+1)(\lambda-\lambda_{0})}}. 
\end{equation} 
These results allow us to write the evolving geometry proposed in (\ref{eq3}) as 
\begin{equation}\label{eq24} 
d{s^2_A} = l^2 \qty[- r^{\pm D \alpha(\lambda,D)} d{t^2}+\frac{1}{r^2} d{r^2}+ r^{\pm  \alpha(\lambda,D)} d{x_{i}}d{x^{i}}], \ \ \ \ \   i = 1,2,\ldots D. 
\end{equation} 
We can see that (\ref{eq24}) is well defined in the interval $\lambda$ $\in$ $]\lambda_{0},\infty[$. Nevertheless, here there is no analytical continuation like the one that takes place in \cite{Arturo} for maximally symmetric spacetimes. We can write (\ref{eq24}) in a canonical Lifshitz form by using the following transformation of coordinates 
\begin{equation}\label{eq25} 
t \to \frac{2}{\alpha}\tilde{t}, \hspace{1.5cm} r \to \tilde{r}^{\frac{2}{\alpha}},\hspace{1.5cm} x_{i} \to \frac{2}{\alpha}\tilde{x}_{i}, 
\end{equation} 
so that the metric adopts the form 
\begin{equation}\label{eq26} 
d{\tilde{s}^2_A} = \tilde{l}^2(\lambda, D)\qty[- \tilde{r}^{\pm 2 D } d{\tilde{t}^2}+\tilde{r}^{-2} d{\tilde{r}^2}+ \tilde{r}^{\pm 2} d{\tilde{x}_{i}}d{\tilde{x}^{i}}], \ \ \ \ \   i = 1,2,\ldots D, \end{equation} with \begin{equation}\label{eq27} \tilde{l}(\lambda,D) = \frac{2 l}{\alpha (\lambda, D)}. 
\end{equation} 
%%
%One of the main results of this work is given by (\ref{eq26}). 
The factor that multiplies the brackets is a constant for a given specific value of the flow parameter $\lambda$. In the limit when $\alpha \to 2$ we identify the spacetime generated in (\ref{eq26}) as the Lifshitz space for the case in which the critical exponent is equal to the spatial dimension. Additionally, the curvature scalar of this geometry is given by 
\begin{equation}\label{eq28} R = -\frac{(5D+1)}{2(D+1)}\qty(\frac{1}{\lambda-\lambda_{0}}). 
\end{equation} 
This invariant is negative definite, and when $\lambda \to \infty$ it goes to zero. This means that in this limit we reproduce the flat spacetime. By considering the definition of fixed points 
\begin{equation}\label{fp} 
\partial_{\lambda} g_{\mu\nu}=0
 \end{equation} 
 applied to the metric (\ref{eq24}) we find the conditions 
 \begin{align} 
 \partial_{\lambda} g_{tt} &= \pm \frac{l^4}{(D+1)(\lambda-\lambda_0)^2} r^{\pm D \alpha (\lambda,D)-1}=0, \label{eq29} \\ 
 \partial_{\lambda} g_{rr} &= 0,\label{eq30}\\ 
 \partial_{\lambda} g_{ii} &= \mp \frac{ l^4}{D(D+1)(\lambda-\lambda_0)^2} r^{\pm \alpha (\lambda,D)-1}=0,\label{eq31} 
 \end{align}
so, in order to satisfy them, we find that $\lambda \to \infty$ for every $r$. This implies that there is a single fixed point along the flow which coincides with the flat spacetime. 
%%%%%%%%% SUBSECTION %%%%%%%%%%%
\subsection{Solutions to the flow with continuous critical exponent} %%%%%%%%
%%%%%%%%%%%
We already know the suitable ansatz to DeTurck vector field $V_{r}(\lambda,r)$ that will allow us to reproduce Lifshitz spaces, so we can obtain a more general solution taking the form of $V_{r,\lambda}$ and noting that is posible to find a more general solution to (\ref{eq12}) that it is not necessarily a constant.
So, we first postulate $V_{r} (r,\lambda)$ as
\begin{equation}\label{postulate}
V_{r}(r,\lambda)  = -\rho (\lambda) \frac{\ln r}{r} +\frac{F(\lambda)}{r}
\end{equation}
a new solution to (\ref{eq12}) is given by 
\begin{equation}\label{w_new}
w(r,\lambda) = \pm\chi(\lambda)  \ln (r) + G(\lambda) 
\end{equation}
provided that 
\begin{equation}
F(\lambda) = - \chi(\lambda) \and \qquad G(\lambda) = \text{const.}=c
\end{equation}
since $w(r,\lambda) =u_{1}(r,\lambda)+D u3(r,\lambda)$, we can choose $u_{1}(r,\lambda)$, $u_{3}(r,\lambda)$ and $F(\lambda)$ as follows
\begin{equation}\label{sol2} 
u_{1}(r,\lambda) = \pm 2 \beta (\lambda)\ln r,\qquad u_{3}(r,\lambda) = \pm 2 \gamma (\lambda) \ln r, \qquad F(\lambda) = -(\beta(\lambda)+D \gamma(\lambda))
\end{equation} 
by substituting these solutions  and  (\ref{postulate}) in (\ref{eq9}), (\ref{eq10}) and (\ref{eq11}) we obtain 
\begin{equation}\label{rel} 
\begin{split} l^{2} \dot{\beta}(\lambda) +\beta(\lambda) \rho (\lambda) &= 0\\ 
l^2 \dot{\gamma}(\lambda)+\gamma(\lambda) \rho (\lambda) &= 0\\ 
\beta^2 (\lambda) + D \gamma ^2(\lambda) - \rho (\lambda) &=0 
\end{split} 
\end{equation}
with solutions given by
\begin{equation}\label{rel2}
\beta(\lambda)= k \sqrt{\frac{l^2}{2(k^2+D) (\lambda-\lambda_{0})}}, \qquad \gamma 
(\lambda) = \sqrt{\frac{l^2}{2(k^2+D) (\lambda-\lambda_{0})}},\qquad 
\rho (\lambda) = \frac{l^2}{2(\lambda-\lambda_{0})}
\end{equation}
were $k$ is a continuous parameter that may have any real value. Then
\begin{equation}\label{f_sols}
f_{1}(r,\lambda) = r^{\pm 2 k \sqrt{\frac{l^2}{2(k^2+D) (\lambda-\lambda_{0})}} }, \qquad 
f_{3} 
(r,\lambda) = r^{\pm 2 \sqrt{\frac{l^2}{2(k^2+D) (\lambda-\lambda_{0})}}}, \qquad F(\lambda) = -(k +D) \sqrt{\frac{l^2}{2(k^2+D) (\lambda-\lambda_{0})}}.
\end{equation}
So the evolving metric has the form
\begin{equation}\label{solB1}
d{s_{B}^2} = l^2 \qty[- r^{\pm 2 k \sqrt{\frac{l^2}{2(k^2+D) (\lambda-\lambda_{0})}} }
d{t^2}+\frac{1}{r^2} d{r^2}+ 
		r^{\pm 2\sqrt{\frac{l^2}{2(k^2+D) (\lambda-\lambda_{0})}}}
d{x_{i}}d{x^{i}}], \ \ \ \ \   i = 1,2,\ldots D.
\end{equation}
if we perform a similar transformation of coordinates like (\ref{eq25}), i. e.:
\begin{equation}\label{coor_trans}
t \to \frac{1}{\gamma (\lambda)}\tilde{t}, \hspace{1.5cm}
r \to \tilde{r}^{\frac{1}{\gamma (\lambda)}},\hspace{1.5cm}
x_{i} \to \frac{1}{\gamma (\lambda)}\tilde{x}_{i},
\end{equation}
the metric adopts the form 
\begin{equation}\label{solB2}
d{\tilde{s}_{B}^2} = \tilde{l}^2(\lambda) \qty[- \tilde{r}^{\pm 2 z} d{t^2}+\frac{1}{\tilde{r}^2} d{\tilde{r}^2}+ 
		\tilde{r}^{\pm2} d{\tilde{x_{i}}}d{\tilde{x}^{i}}], \ \ \ \ \   i = 1,2,\ldots 
D,
\end{equation}
with 
\begin{equation}\label{l2}
\tilde{l}^2(\lambda) = \frac{2 (z^2+D) (\lambda-\lambda_0)}{l^2},
\end{equation}
is easy to note that the solution (\ref{solB2}) is more general than 
(\ref{eq26}), this follows from the fact that now the 
critical exponent ($z=k$) may adopt continuous  values since is an arbitrary continuous constant.  
%In Fig. \ref{zplot} we see the dependence between the critical exponent and the values that may adopt the constant $k$, we can observe that when $k$ closes to $l^2$, the critical exponent becomes much and much larger. 
%
%\begin{figure}[h!]
%\centering
%\includegraphics[scale=1]{z.eps}
%\caption{The critical parameter $z$ as a function of $k$ with $l=10$ and $d=3$. }
%\label{zplot}
%\end{figure}
 %
 The curvature scalar in the continuous case is given by
 \begin{equation}\label{scalar2}
 R = -\qty(\frac{D+D^2+2 z D+2 z^2}{2(D+z^2)}) \frac{1}{\qty(\lambda-\lambda_0)},
 \end{equation}
Following similar arguments like that in the discrete case, is easy to check that this flow has one fixed point that also coincides with the flat spacetime. 

\section{Discussion and concluding remarks}\label{s3}
The solutions that we have obtained from the evolution of the Hamilton-De Turck Ricci flow (\ref{eq2}) give us a family of Lifshitz spaces with both discrete and continuous critical exponents. The latter solution is more general than the former one, since in it the critical exponent may adopt any real value including those corresponding to the discrete case. %However, it was convenient to place both results, since the discrete case allowed us to find the suitable aspect of the DeTurck vector field.
Thus we have shown that Lifshitz spacetimes are exact solutions of the Hamilton-DeTurck Ricci flow equations when starting with a general setup given by (\ref{eq3}). Therefore, Lifshitz geometries can be constructed in a purely geometric way with the aid of the Ricci flow, in contrast to the RG flow formalism that makes use of several $\sigma$-models defined in terms of diverse fields. In this sense, we are generating Lifshitz spaces in an universal way since these geometries can be supported by several actions where gravity is coupled to a wide range of matter fields \cite{Taylor,HLS}. Additionally, we see that the spaces that we have behave in a similar way to those discussed in \cite{Oli}, \textit{i. e.}, they preserve the property that the flow is well defined for the interval $\lambda$ $\in$ $]\lambda_0,\infty[$, and in the limit when $\lambda \to \infty$, we recover the flat spacetime geometry; we see this fact in an invariant way from the form of the curvature scalar. With this in mind, we can say that, analogously to what happens with the diffusion of heat, for large enough values of the flow parameter, the Ricci flow equations homogenize the metric in such a way that it reaches the equilibrium state given by the flat spacetime. 

We would like to contrast the result regarding the fixed point for both the obtained metric (\ref{eq24}) and the transformed  interval (\ref{eq26}). For the former, the fixed point corresponds in a clear way to flat spacetime as $\lambda \to \infty$, which is in agreement with the behavior of the curvature scalar when the Ricci flow parameter adopts large values. By following the definition of a fixed point (\ref{fp}) for the transformed metric (\ref{eq26}), it is clear that we also need to transform the flow parameter 
\begin{equation}\label{eq32}
\lambda \to \tilde{\lambda}(\lambda) 
\end{equation}
in such a way that the partial derivative respect to the Ricci flow parameter now reads 
\begin{equation}\label{eq33} 
\partial_{\lambda} \to \pdv{\tilde{\lambda}}{\lambda} \partial_{\tilde{\lambda}} 
\end{equation} 
where $\tilde{\lambda}$ is a function of $\lambda$, that we shall call $P(\lambda)$.  With this transformation we can find the fixed point equations for the metric in question, leading to the following relations
 \begin{align} 
 \partial_{\tilde{\lambda}} \tilde{g}_{tt}=0 & \to -\frac{2 D(D+1)}{\partial_{\lambda} P }\tilde{r}^{\pm 2D}=0, \label{eq34}\\ 
 \partial_{\tilde{\lambda}} \tilde{g}_{rr} =0 & \to  \frac{2 D (D+1)}{\partial_{\lambda} P} \tilde{r}^{-2}=0,\label{eq35}\\ 
 \partial_{\tilde{\lambda}} \tilde{g}_{i i} =0 & \to   \frac{2 D (D+1)}{\partial_{\lambda} P} \tilde{r}^{\pm 2} =0 ,\label{eq36} 
 \end{align} 
therefore, in order to recover the fixed point already computed for the original metric (\ref{eq24}), we have to impose the condition that $\partial_{\lambda} P(\lambda) \to \infty$ as $\lambda \to \infty$; a good choice of the function $P$ would be, for example, $P(\lambda) \sim \lambda^{n}$ with $n > 1$. On the other hand,  with this choice we also recover the behavior of the curvature scalar (\ref{eq28}) as $\tilde{\lambda} \to \infty$, \textit{i. e.} we obtain the flat spacetime as a fixed point along the flow.  

A similar analysis of the metric with a continuous critical exponent (\ref{solB2}) leads us to the same results regarding the fixed points of the Ricci flow: the family of geometries evolves towards flat spacetime. 

It would be interesting as well to consider the evolution of spatially anisotropic\footnote{Those spaces with translational symmetry in which each direction scales differently.} but homogeneous Lifshitz geometries  under the Hamilton-DeTurck Ricci flow like the spaces recently constructed in \cite{Taylor,Viri}, \textit{i. e.} when the Lifshitz geometry admits a non relativistic scaling symmetry given by
\begin{align}\label{eq:3}
D_{z}:\ \ & r \longrightarrow r'=\lambda^{\pm 1}r\nonumber \,, \\
& t \longrightarrow t'=\lambda^{\mp z}t, \\
& x^{i}\longrightarrow x^{'i}=\lambda^{\mp z_{i}} x^{i} \,. 
\nonumber
\end{align}
where $z_{i}$ are critical exponents. As a final remark, we would like to point out that the above constructed family of Lifshitz spaces includes two relevant cases for $D=z=2$ and $D=z=3$ since within the Gravity/Condensed Matter correspondence they describe physically relevant quantum systems when the critical exponent is $z=2$ for two-dimensional materials, and $z=3$ for three-dimensional solid state systems \cite{HLS}.
		
\section*{Acknowledgments} RCF and AHA acknowledge support from SNI-CONACyT and VIEP-BUAP. JAHM is grateful to CONACyT for support.

    %
    % For  figures use
    %\begin{figure*}
    %\vspace*{5cm}       % Give the correct figure height in cm
    %\includegraphics{leer.eps}
    %\end{figure*}
    % or  this
    %\begin{figure}
    %\centering
    %\resizebox{0.75\textwidth}{!}{%
    %  \includegraphics{leer.eps}
    %\vspace{5cm}       % Give the correct figure height in cm
    %\end{figure}
    %\begin{table}
    %\centering
    %\begin{tabular}{lll}
    %\hline\noalign{\smallskip}
    %\end{tabular}
    % Or use
    %\vspace*{5cm}  % with the correct table height
    %\end{table}

\begin{thebibliography}{10} 
\bibitem{Hamilton} R. S. Hamilton, J. Diff. Geo. \textbf{17}, 255-306 (1982). 
\bibitem{Turck} D. M. DeTurck, J. Diff. Geo. \textbf{18}, 157-162 (1983). 
\bibitem{Perelman1} G. Perelman, \textit{The entropy formula for the Ricci flow and its geometric applications,} \href{https://arxiv.org/abs/math/0211159v1}{arXiv:math/0211159v1 [math.DG]}, (2002). 
\bibitem{Perelman2} G. Perelman, \textit{Ricci flow with surgery on three-manifolds,} \href{https://arxiv.org/abs/math/0303109v1}{ arXiv:math/0303109v1 [math.DG]}, (2003). 
\bibitem{Husain} V. Husain and S. S. Seahra, Class. Quantum Grav. \textbf{25}, 222002 (2008). 
\bibitem{Dai} X. Dai and L. Ma, Commun. Math. Phys. \textbf{274}, 65 (2007). 
\bibitem{Woolgar2012} T. Balehowsky, E. Woolgar, J. of Math. Phys. \textbf{53}, 072501 (2012). 
\bibitem{Samuel2007} J. Samuel, S. R. Chowdhury, Class. Quantum Grav. \textbf{24}, F47 (2007). 
\bibitem{Samuel2008} J. Samuel, S. R. Chowdhury, Class. Quantum Grav. \textbf{25}, 035012 (2008). 
\bibitem{Wiseman} M. Headrick and T. Wiseman, %\textit{Ricci flow and black holes}, 
Class. Quantum Grav. \textbf{23}  66836707 (2006). 
\bibitem{Woolgar2008} E. Woolgar, Can. J. Phys. \textbf{86}, 645 (2008). 
%\bibitem{Woolgar1} E. Woolgar, \textit{Some Applications of  Ricci Flow in Physics,} \url{https://arxiv.org/abs/0708.2144}, (2003). 
\bibitem{jackson} S. Jackson, R. Pourhasan and H. Verlinde, \textit{Geometric RG Flow,} arXiv:1312.6914 [hep-th]. 
\bibitem{kiritsis} E. Kiritsis, W. Li and F. Nitti,  Fortsch. Phys. \textbf{62}, 389 (2014) doi:10.1002/prop.201400007 [arXiv:1401.0888 [hep-th]]. 
\bibitem{horava} P. Ho\v{r}ava, JHEP \textbf{0903}, 020 (2009) doi:10.1088/1126- 6708/2009/03/020 [arXiv:0812.4287 [hep-th]]. 
\bibitem{nishioka} T. Nishioka, Class. Quant. Grav. \textbf{26}, 242001 (2009) doi:10.1088/0264- 9381/26/24/242001 [arXiv:0905.0473 [hep-th]]. 
\bibitem{Figueras} P. Figueras, J. Lucietti and T. Wiseman, Class. Quantum Grav. \textbf{28}, 215018 (2011). 
\bibitem{Arturo} R. Cartas-Fuentevilla, A. Herrera-Aguilar, J. A. Olvera-Santamar\'ia, Eur. Phys. J. Plus \textbf{133}, 235 (2018). 
\bibitem{Taylor} M. Taylor, Class. Quantum Grav. \textbf{33}  033001, (2016). 
\bibitem{Viri} R. Cartas-Fuentevilla, A. Herrera-Aguilar, V. Matlalcuatzi-Zamora, U. Noriega and J. Romero, \textit{Anisotropic Lifshitz holography in Einstein-Proca theory with stable negative mass spectrum,} \href{https://arxiv.org/abs/1804.02278v1}{arXiv:1804.02278v1[hep-th]}, (2018).
\bibitem{Kachru} S. Kachru, X. Liu and M. Mulligan,
%\textit{Gravity duals of Lifshitz-like fixed points,} 
Phys. Rev. D \textbf{78} 106005 (2008).
\bibitem{HLS} S. A. Hartnoll, A. Lucas and S. Sachdev, \textit{Holographic quantum matter,} \href{https://arxiv.org/abs/1612.07324v3}{arXiv:1612.07324v3 [hep-th]}, (2018). 
 
\bibitem{Topping} P. Topping, \textit{Lectures on the Ricci flow}, Cambridge University Press, (2006). 
\bibitem{benneth} B. Chow, D. Knopf, \textit{The Ricci flow: An Introduction}, Mathematical Surveys and Monographs \textbf{110}, (2004). 
\bibitem{Oli} T. Oliynyk, E. Woolgar, Commun. Anal. Geom. \textbf{15}, 535 (2007). 

\end{thebibliography}
    \end{document}